# Observations of the fluctuations in Interplanetary Magnetic Field around L1 point during the solar transient events of 2024 with MAG payload onboard Aditya-L1 spacecraft

Vipin K. Yadav[1], R.K. Choudhary[1] and M. Midhun Goutham[2]

1 Space Physics Laboratory (SPL), Vikram Sarabhai Space Centre (VSSC), ISRO, Thiruvananthapuram 695022, Kerala, India

2 Sardar Vallabhbhai National Institute of Technology, Ichchhanath, Surat 395007, Gujarat, India

Email

Vipin K. Yadav: vipin_ky@vssc.gov.in; R.K. Choudhary: rajkumar_choudhary@vssc.gov.in

ORCID

Vipin K. Yadav: 0000-0002-1470-8443; R.K. Choudhary: 0000-0002-1276-0088

**Abstract**

The MAG payload onboard India's first solar mission, Aditya-L1, is a dual-sensor fluxgate magnetometer designed to measure the interplanetary magnetic field (IMF) while operating in a halo orbit around the first Sun-Earth Lagrangian point (L1). Since becoming operational in January 2024, MAG has continuously recorded local magnetic field data and has captured several solar transient events over the past one year. During these events, the IMF, typically around 5 nT, exhibited significant enhancements in magnitude. This study focuses on three such solar events observed in March, May, and October 2024. Analysis of the magnetic field power spectra during these events reveals fluctuations consistent with Kolmogorov-type turbulence, characterized by a spectral slope close to $-5/3$. To emphasize changes in spectral behavior, the event-day spectra are compared with those from a day when the quiet solar wind conditions prevail. A marked contrast is observed: while the quiet periods exhibit anisotropic turbulence, the extreme events display quasi-isotropic behavior, with spectral slopes closely following the Kolmogorov spectrum across all three IMF components. The results, including detailed variations in spectral slope and turbulence characteristics, are presented and discussed in this paper.

## 1. Introduction

The Sun continuously emits energy and matter into the heliosphere in the form of three primary physical parameters, namely (a) the electromagnetic radiation; (b) charged particles; and (c) the magnetic field. The radiation spans across various wavelengths, including infrared (IR), visible, and ultraviolet (UV) light. The charged particle component primarily consists of electrons, protons, and heavier ions, collectively known as solar wind. Accompanying the solar wind is the Sun's magnetic field, which is carried into interplanetary space as the interplanetary magnetic field (IMF). The IMF is not emitted independently but is embedded within and transported by the solar wind, making the two inseparable.

At a distance of approximately 1 astronomical unit (AU) from the Sun near the Earth's orbit, the typical strength of the IMF is about 5 nT (Wang & Sheeley, 1988; Zhao & Hoeksema, 1995). However, this IMF magnitude is not constant; both the magnitude and the direction of IMF undergo significant variability, particularly during periods of intense solar activity such as solar flares and coronal mass ejections (CMEs) (Wang, 2003). These temporal and spatial variations are broadly referred to as magnetic fluctuations. Such fluctuations occur across a wide range of scales and frequencies, reflecting the inherently dynamic and turbulent nature of the solar wind.

The origin of these magnetic fluctuations can be traced back to the turbulent motions in the solar corona, from where the solar wind is believed to be generated. These fluctuations propagate through the heliosphere and are governed by complex plasma processes. In general, due to the anisotropic and nonlinear characteristics of the solar wind plasma, the resulting magnetic field fluctuations exhibit turbulent behaviour, with energy cascading from larger to smaller scales, a hallmark of magnetohydrodynamic (MHD) turbulence (Kawashima, 1969). These fluctuations are often intermittent, with varying amplitudes and durations, and can appear as isolated bursts or as part of large-scale coherent structures (Bruno, 2007).

It is important to note that the study of IMF fluctuations has direct implications for space weather forecasting. Of particular interest is the north-south component of the IMF, commonly denoted as Bz. When oriented southward, a negative Bz can strongly couple with Earth's geomagnetic field, enabling the transfer of energy and momentum into the magnetosphere. This process can initiate a range of geophysical phenomena, including auroral intensification, geomagnetic storms, and ionospheric disturbances (Garrett, 1974).

To investigate these magnetic fluctuations, in-situ measurements obtained by spacecraft positioned at strategic locations throughout the heliosphere have been extensively utilized (Couzens & Kings, 1986; Hapgood et al., 1991; Smith & Lockwood, 1996). Magnetic field measuring instruments such as the magnetometers provide high-resolution data on the IMF, enabling the construction of power spectra to identify dominant frequencies and analyze the scale-dependent distribution of magnetic energy. Additionally, theoretical and computational models based on MHD turbulence theory are employed to interpret the observational data and gain insight into the underlying plasma dynamics (Wang & Sheeley, 1988; Hairston & Heelis, 1990, Zhao & Hoeksema, 1995: Stamper et al., 1999).

The present study aims to investigate the magnetic fluctuations in IMF using the MAG payload measurements onboard Aditya-L1 spacecraft around L1 point, a stable vantage point for solar observations, during the solar transient events of 2024. The paper is structured as follows: Section 2 provides a detailed overview of the nature and characteristics of magnetic fluctuations in the IMF. Section 3 presents observational data obtained from Aditya-L1. Section 4 highlights notable extreme solar events that occurred in March, May, and October 2024. An analysis of the observed magnetic fluctuations based on spectral and statistical methods is presented in section 5. Section 6 concludes with a discussion of the implications of these findings for space weather studies and future research directions.

**2. Magnetic Field Fluctuations and Anisotropy in the IMF**

In interplanetary space, the magnetic field fluctuations, particularly in the IMF, exhibit anisotropy in the steady state. Typically, fluctuations perpendicular to the magnetic field lines are much larger than those along the magnetic field lines. However, this anisotropy diminishes as the fluctuation amplitude increases, eventually becoming isotropic (Kawashima, 1969). These IMF fluctuations are also linked to solar wind ion thermal anisotropy. At higher fluctuation amplitudes, beyond a certain threshold, nonlinear effects can convert transverse fluctuations into longitudinal ones. Additionally, the anisotropy is frequency dependent: at lower frequencies, it tends to become more isotropic (Kawashima, 1969).

The anisotropy observed in the IMF magnetic fluctuations correlates with the behavior of solar wind protons in the framework of MHD turbulence. In regimes of low turbulent Mach numbers, the compressive component of plasma appears to be driven by the incompressible turbulent flow, suggesting that local fluid dynamics primarily govern the compressive behavior (Smith, 2006).

It is also proposed that the IMF anisotropy is strongly influenced by local plasma conditions, particularly the proton plasma beta - the ratio of kinetic to magnetic pressure (Smith, 2006). Investigations into the scaling behavior of the IMF magnitude in relation to variations in solar wind velocity fields have revealed similarities with the scaling of passive scalars in hydrodynamic turbulence, especially during solar maximum periods. This indicates that the scaling properties of the IMF magnitude vary over the solar cycle. Furthermore, magnetic field magnitude fluctuations have been observed to be significantly more intermittent than velocity fluctuations in both fast and slow solar winds (Bruno, 2007).

The power spectral densities of magnetic field fluctuations associated with interplanetary shocks often exhibit Alfvénic characteristics. These fluctuations are also linked to energetic particle events, with particle energy spectra reflecting the power spectra of associated magnetic field fluctuations (Qiang, 2013). Fast interplanetary coronal mass ejections (ICMEs) are preceded by sheaths, the highly turbulent and fluctuating regions. In these sheaths, magnetic field fluctuations are characterized by increased power and reduced anisotropy, especially when fast magnetic clouds interact with an already turbulent solar wind (Moissard, 2019). These regions also display large-amplitude magnetic field fluctuations that evolve as the ICME propagates from the Sun to the Earth (Good, 2020).

Lower magnetic fluctuation amplitudes are typically observed in radial IMF intervals, across both MHD and kinetic scales. During these intervals, the proton temperature tends to be more isotropic, and there is a higher occurrence of plasma waves in the 0.1–1 Hz frequency range (Pi, 2022). As the hot solar wind and strong magnetic fields from the solar corona expand through the heliosphere, observations from the Parker Solar Probe (PSP) have provided key insights. Specifically, PSP measurements below the Alfvén critical surface, approximately 13 million kilometres from the Sun's photosphere, revealed a magnetically dominated region with very low plasma beta values (Kasper, 2021), with

$$\beta = \frac{nK_B T}{B^2/\mu_0} \ll 1$$

Beyond magnetic field fluctuations, plasma turbulence and flow speeds in the solar corona have also been investigated using radio-sounding experiments. In these studies, Doppler-shifted radio signals received on Earth are spectrally analyzed to determine Doppler residuals and, in turn, the turbulence spectrum (Jain et al., 2023). Similarly, spectral analysis near coronal holes has been employed to study electron density fluctuations and flow speeds, offering insights into how the solar wind gains energy as it traverses these regions (Jain et al., 2024; Aggarwal et al., 2025).

2.1 Kolmogorov Scaling in Magnetic Field Fluctuations

When the energy density of magnetic field fluctuations scales with the wave number $k$ as $k^{-5/3}$, it indicates that the power-law distribution follows the Kolmogorov magnetic field spectrum. This behaviour is characteristic of a turbulent magnetic field, where the fluctuation energy cascades across scales in accordance with Kolmogorov's theory of fluid turbulence. The presence of this scaling within the inertial range is a distinct signature of fully developed turbulence (Kolmogorov, 1991, Xu, 2023).

Similarly, in the solar wind, magnetic field fluctuations exhibit turbulent behavior, characterized by a magnetic energy spectrum with a power-law index of -5/3, consistent with the Kolmogorov spectrum (Lee, 2020, Jain, 2023, Jain, 2024). A notable feature of this spectrum in the inertial range is the transfer of energy from larger to smaller scales without appreciable dissipation. This Kolmogorov-type scaling is frequently studied within the framework of MHD, which describes the interactions between magnetic fields and plasma flows.

Direct, *in-situ* observations of magnetic turbulence in the interstellar medium (ISM) were made by the *Voyager 1* spacecraft between 2012 and 2019, when it was at distances up to 26 AU beyond the heliopause. The magnetic power spectrum of fluctuations in the ISM followed a Kolmogorov power law, exhibiting a spectral index of (-5/3). Furthermore, the power in perpendicular magnetic fluctuations was generally observed to be greater than that in parallel fluctuations, indicating the presence of hydromagnetic waves in the turbulent spectrum. Alongside magnetic field fluctuations, electron density fluctuations in the inertial range also followed a Kolmogorov scaling law, with a one-dimensional power-law index of (-5/3) (Lee2020).

Similar Kolmogorov-type scaling in magnetic field fluctuations has been observed in Saturn's magnetosphere. Data from instruments aboard the *Cassini* spacecraft revealed the presence of such scaling in the inertial range across all local times -including the noon-side, dusk-side, dawn-side, and night-side - indicating the widespread presence of turbulence in planetary magnetospheres (Xu et al., 2023).

### 3. MAG Observations of the IMF around L1 Point

The Aditya-L1 mission, India's first dedicated solar observatory, is currently operating in a halo orbit around the first Lagrange (L1) point, enabling continuous observation of the Sun (Tripathi, 2022). The mission is equipped with a suite of remote sensing instruments that cover multiple wavelengths, namely the Visible Emission Line Coronagraph (VELC) (Singh et al., 2025), Solar Ultraviolet Imaging Telescope (SUIT) (Tripathi et al., 2025), and two X-ray payloads - HEL1OS and SoLEXS (Sankar et al., 2025) to study the Sun across the electromagnetic spectrum. Additionally, three *in-situ* payloads are on board: the Aditya Solar Wind Particle Experiment (ASPEX) (Goyal et al., 2025; Kumar et al., 2025), which measures energetic ions and alpha particles in the solar wind; the Plasma Analyser Package for Aditya (PAPA) (Thampi et al., 2025), which analyzes low-energy ions and electrons; and the fluxgate magnetometer (MAG), designed to monitor the local interplanetary magnetic field (IMF) and its variations during extreme solar transient events (Janardhan et al., 2017).

The primary scientific goals of the MAG payload include detecting interplanetary coronal mass ejections (ICMEs), investigating the impact of solar transients on near-Earth space weather, and identifying signatures of solar plasma waves near the L1 point (Yadav et al., 2018). MAG comprises two identical triaxial sensor units mounted on a 6 m long deployable boom, with one sensor positioned at the tip (6 m from the spacecraft) and the other midway (3 m from the spacecraft). This configuration enables differential measurements that effectively cancel out spacecraft-generated magnetic fields, enhancing the accuracy of IMF observations (Yadav et al., 2025).

Since commencing operations, MAG has recorded several significant solar transients, including coronal mass ejections (CMEs) and magnetic clouds, typically characterized by pronounced enhancements in the interplanetary magnetic field (IMF) from its nominal value of ~ 5 nT (Wang & Sheeley, 1988). One such intense event occurred on 22-24 March 2024, when the IMF peaked at 38 nT with Bz fluctuations of approximately ±26 nT. An even stronger disturbance was observed on 10-11 May 2024, when the IMF reached 82 nT and the Bz component varied by nearly ±70 nT. These disturbances were traced to multiple CMEs erupting on 8 May, one of which impacted the L1 point on 10 May and subsequently triggered a severe geomagnetic storm upon arrival at Earth (A. Jain et al., 2025, Ambili et al., 2025). Another notable event was detected on 10-11 October 2024, when a magnetic cloud passed through L1, exhibiting a rotating Bz component with extreme values of ~ ±40 nT. The following sections detail these transients and their signatures in solar wind magnetic field fluctuations.

### 4. Extreme Solar Transient Events

Three extreme solar transient events are analyzed in this study, one each in the months of March, May and October, 2024.

4.1 March 2024 Event: A case of "strong" geomagnetic storm

An extreme solar transient event occurred between March 23 and 24, 2024, leading to a significant geomagnetic disturbance. The minimum Dst (Disturbance Storm Time) index recorded during this event dropped to approximately -120 nT by 16:00 UT on March 24, indicating the occurrence of a strong geomagnetic storm. According to the classification by (Loewe, 1997), geomagnetic storms are categorized based on their minimum Dst values: weak storms (-30 to -50 nT), moderate storms (–50 to –100 nT), strong storms (-100 to -200 nT), severe storms (-200 to -350 nT), and great magnetic storms for values below -350 nT. An example of such a great storm occurred on May 10-11, 2024.

Figure 1 illustrates the evolution of IMF components - Bx, By, Bz and |B|, the total IMF magnitude, as measured by the MAG instrument around the L1 point during this period in Geocentric Solar Ecliptic (GSE) coordinates. To provide context, we have included the data beginning from March 22 to highlight IMF behavior during quiet conditions. The magnetic field remained relatively weak until 00:00 UT on March 23, which corresponded with low Dst values at the ground (lower panel of Figure 1).

Subsequently, mild fluctuations in all three IMF components were observed, with Dst values staying below -100 nT.

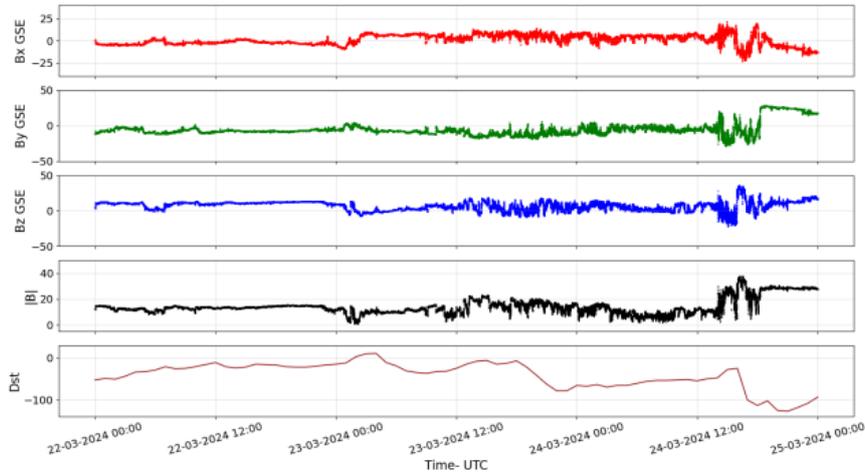

Figure 1: IMF components Bx, By, Bz and total |B| magnitude recorded by MAG around L1 point during March 22–24, 2024 in GSE coordinates. The lower panel shows the corresponding Dst index measured at the ground level. The quiet IMF conditions prior to March 23 and the subsequent increase in fluctuations - especially after 14:00 UT on March 24 -highlight the onset and progression of a strong geomagnetic storm.

After 14:00 UT on March 24, intense variations in the IMF components emerged, culminating in a strong geomagnetic storm in near-Earth space. During this interval, the IMF magnitude fluctuated sharply, reaching up to ±30 nT. The maximum values of the total IMF and its components during the March 22-24, 2024 event are summarized in Table 1.

Table 1: Maximum values of the total IMF and its components (Bx, By, Bz) recorded during the solar transient event of March 22–24, 2024.

| IMF Variation | Minimum (in nT) | Maximum (in nT) |
| --- | --- | --- |
| Bx | -21 | +20 |
| By | -21 | +36 |
| Bz | -28 | +26 |
| |B| | -- | +38 |

As shown in the Figure 1, the onset of geomagnetic storm is clearly evident through significant fluctuations in the IMF parameters, which become more pronounced following the sharp decline in Dst values. The Bz component of the IMF, critical for geomagnetic coupling, reaches a minimum of approximately –28 nT, while the total magnetic field strength peaks around +38 nT. It is also notable that there was minimal variation in the IMF on March 22, a day characterized by quiet solar conditions. To provide a reference for quiet-time behavior, we also present in Figure 2 the IMF variations observed on March 02, 2024, which was the quietest day of the month. On that day, the total magnetic field magnitude remained within the range of 4 to 12 nT.

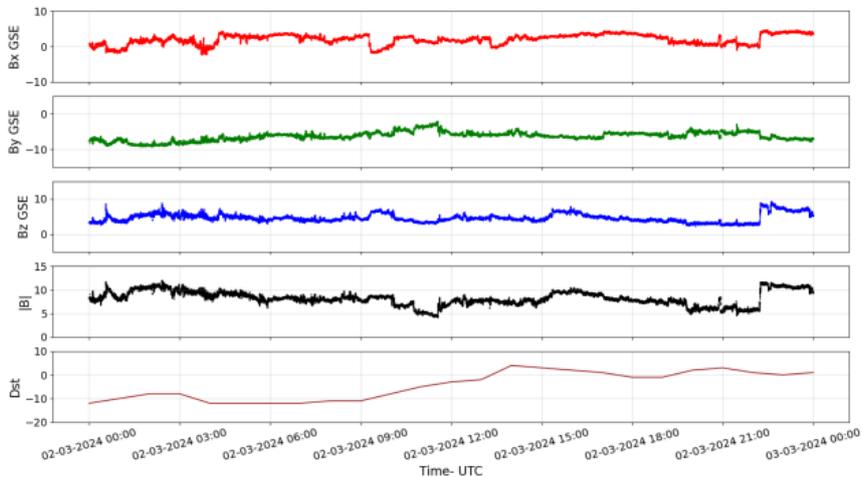

Figure 2: IMF components Bx, By, Bz and total |B| recorded by MAG around the L1 point during March 02, 2024, a relatively quietest solar active day of the month.

4.2 May 2024 Event: A case of "great" geo-magnetic storm

An extreme solar transient event occurred between May 10 and 12, 2024, during which Dst values dropped below -400 nT, indicating a great geomagnetic storm. This event attracted significant attention from the space science community, prompting several studies focused on understanding its geo-effectiveness and its impacts on the near-Earth space environment (Ram et al., 2024; Hayakawa et al., 2025; Ambili et al., 2025).

In Figure 3, we present the variations in the IMF parameters recorded at the L1 point by MAG during this period. The corresponding Dst index fluctuations are shown in the bottom panel.

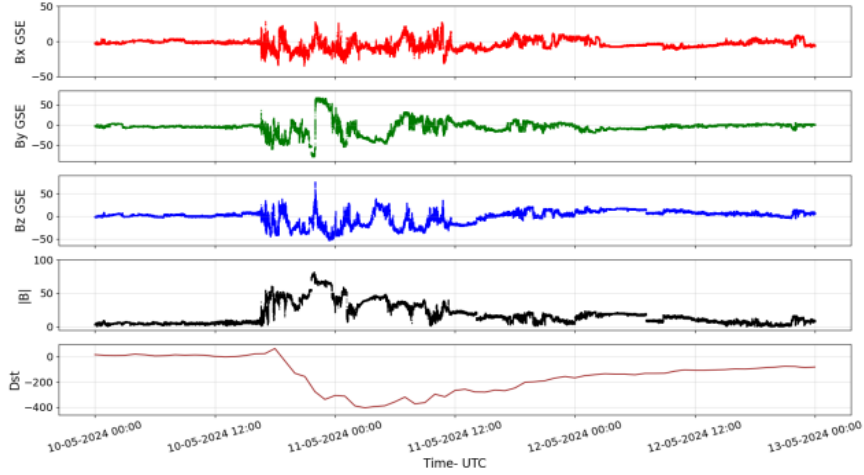

Figure 3: Same as Figure 1 but for May 10-12, 2024 extreme solar event.

It is to be noted from Figure 3 that the magnitude of IMF parameters had sharp jump mapping the dip in the Dst with IMF Bz values fluctuating between ±50 nT, and total |B| elevating crossing 80 nT. Fluctuations in the IMF parameters lasted for over 30 Hours (between 18:00 UT on 10 May to 00 UT on May 12), though the Dst values took longer time to attain the normal quiet-time values. Table 2 lists the minimum and maximum values of IMF parameters. Compared to the March event of the strong geomagnetic storm, both the fluctuations and magnitude of IMF values were intense. The overall IMF variation was also very high at +82 nT, which severely affect the near-Earth space weather leading to a great geomagnetic storm on the Earth.

Table 2: The IMF variation during May, 2024 solar event.

| IMF Variation | Minimum (in nT) | Maximum (in nT) |
|---|---|---|
| Bx | -30 | +28 |
| By | -52 | +70 |
| Bz | -70 | +63 |
| |B| | -- | +82 |

4.3 October 2024 Event: A case of "severe" geo-magnetic storm

A case of "severe" geomagnetic storm occurred during October 10-11, 2024 with a minimum Dst value of ~ -250 nT. Variations in the magnitude of IMF parameters along with the Dst values between 00 UT on October 10 and 24:00 UT on October 11,0 2024 are shown in Figure 4. Apart from sharp fluctuations in IMF parameters, a signature of magnetic cloud with IMF Bz being negative for more than 12 hours during ~ 21:00 UT on October 10 and 09:00 UT on October 11 is quite evident. Maximum and minimum values of IMF parameters (Bx, By, Bz, and |B|) are given in Table 3. It is to be noted that variations in IMF southward component Bz has been moderate (~ -38 nT), compared to great magnetic event on May 10-12, 2024. The overall IMF variation is also moderate at +48 nT, which is higher than the March, 2024 event but lower than the May, 2024 event.

Table 3: The IMF variation during October, 2024 solar event.

| IMF Variation | Minimum (in nT) | Maximum (in nT) |
|---|---|---|
| Bx | -28 | +30 |
| By | -48 | +38 |
| Bz | -38 | +42 |
| |B| | -- | +48 |

5. Fluctuations in the IMF

Variations in the IMF beyond its nominal values indicate excess magnetic fluctuations, often linked to transient solar events. To distinguish these fluctuations from the IMF data recorded by the MAG instrument, we used a method described in (Xu et al., 2023). We employed a one-hour boxcar moving average to extract magnetic field fluctuations (B(t) – [B]), as shown in Figure 5. Here [B] is the one-hour moving average of B(t) used to filter out short-timescale variations. This figure presents variations in the IMF Bx component observed on 24 March 2024 as a test case. The green curve in the top left panel represents raw data, while the black, red, and blue curves correspond to boxcar smoothing with 30-minute, 1-hour, and 2-hour windows respectively. The 1-hour (red) curve best captures the average trend while maintaining relevant variability. Therefore, we applied a consistent 1-hour boxcar moving average throughout our analysis. The bottom left panel of Figure 5 shows the mean-subtracted IMF Bx data.

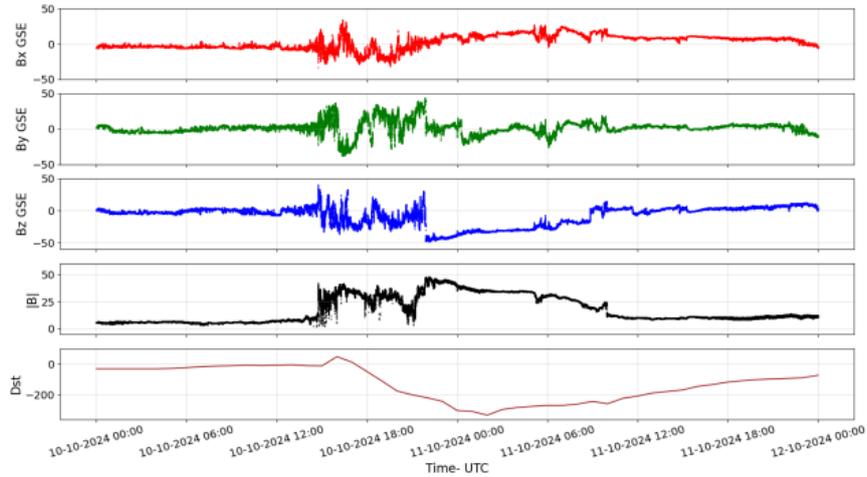

Figure 4: The solar event in October, 2024.

After pre-processing, the IMF time series is transformed into the frequency domain using the Fast Fourier Transform (FFT) to resolve its frequency components and amplitudes. To enhance frequency resolution and produce a smoother spectrum, the $2^n$-point dataset is zero-padded with $2^{n+4} - 2^n$ additional zeros. Frequency bins are computed from the sampling rate using standard FFT routines. The power spectral density (PSD) is then obtained via the Welch method (Welch, 2003) by partitioning the zero-padded series into overlapping segments, each tapered with a Hanning window to suppress spectral leakage. Periodograms are computed from the squared magnitudes of the FFT output for each segment and subsequently averaged. The resulting PSD is normalized to its maximum value to enable comparison across different time scales and solar events. Frequency and PSD are plotted on logarithmic axes to highlight spectral characteristics. As illustrated in the top and bottom right panels of Figure 5, Welch averaging significantly reduces high-frequency fluctuations evident in the unsmoothed PSD.

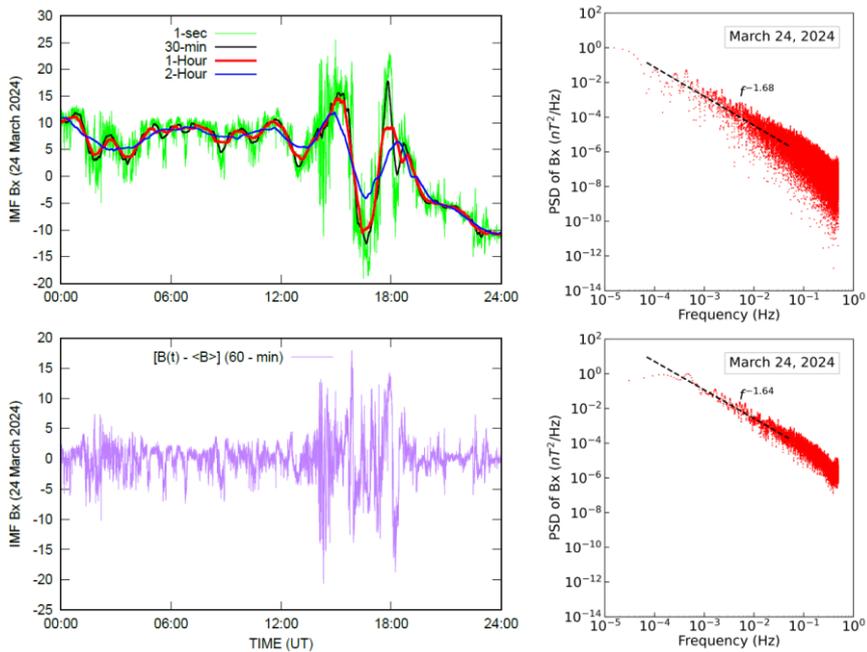

Figure 5: Top left pane: Temporal variations in the IMF Bx during March 24, 2024. The green curve represents 1-sec data, while the black, red, and blue curves correspond to boxcar smoothing with 30-minute, 1-hour, and 2-hour windows, respectively. Bottom left panel: 1-hour box-car mean separated IMF Bx values. Top right panel: PSD of the time series data shown in the bottom left panel. Bottom right panel: Welch smoothed PSD.

To quantify the spectral characteristics, linear regression is performed using the *np.polyfit()* function from Python's NumPy module over a frequency range defined, following (A. Jain et al., 2023), by the length of the processed interval as the lower limit and one-tenth of the Nyquist frequency as the upper limit (Efimov et al., 2008). The slope and intercept from the log--log fit correspond to the power-law relation (the power scaling slope, PSS)

$$\log(y) = k \log(x) + c \quad => \quad y = ax^k$$

where *k* represents the spectral slope.

The analysis is compared with the Kolmogorov scaling law, which characterizes turbulence in magnetic field fluctuations. The right panels of Figure 5 show PSD profiles for the quiet interval (05:00--09:00~UT) on 24 March, derived from the Bx component of IMF fluctuations computed from the 1 s (B(t) – [B]) series. The upper panel corresponds to the unsmoothed spectrum, and the lower to the Welch-smoothed case.

Over the specified frequency range, the estimated slopes are 1.68 and 1.64, respectively, with the latter providing a more stable estimate due to reduced high-frequency fluctuations.

This methodology is applied to investigate magnetic fluctuations during three extreme solar transient events that occurred in 2024. The study utilizes 1-second-averaged IMF data captured by the MAG instrument for high temporal resolution.

5.1 Magnetic field fluctuations during the extreme solar events

The power spectra of the total magnetic fluctuations in the IMF between 14:00 and 18:30 UT (the period when intense fluctuations in IMF B is noted as shown in Figure 1 during March 22-24, 2024 storm event is shown in the top right panel of Figure 6.

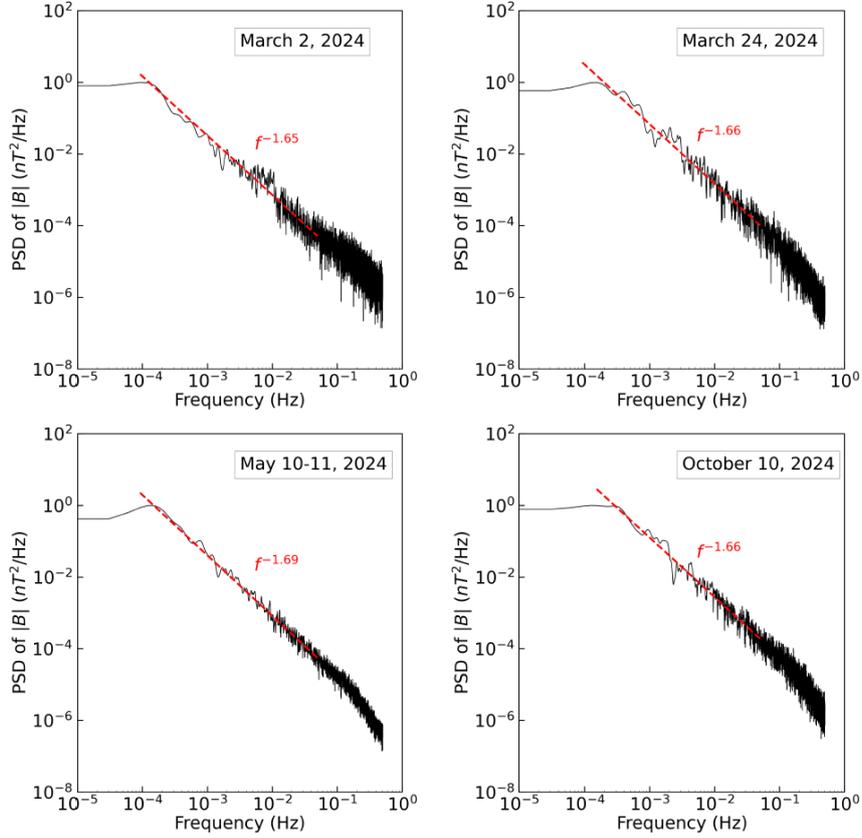

Figure 6: The IMF fluctuations during the extreme solar events are compared with respect to the quiet time conditions. The top right represents the strong geomagnetic storm event day March 24, 2024, while the bottom left and right panels are for the great geomagnetic storm on May 10-11, and the intense storm on October 10, 2024. The top left panel serves the purpose of representing field fluctuations in the solar wind during a geomagnetically quiet day on March 02, 2024.

In Figure 6, the PSD is plotted with the frequencies present in the IMF fluctuations and the PSS is estimated as described above. The PSS value in the total magnetic field fluctuations for the March 24 storm event comes to ~ -1.66. The sampled data is for the period between 14:00 and 18:30 UT on March 24, 2024. It follows the Kolmogorov inertial-range scaling (Kolmogorov, 1991):

$$P(f) \propto f^{-\alpha}, \qquad \alpha \approx \frac{5}{3}$$

indicating fully developed turbulence. Here α represents the PSS magnitude. It is, however, interesting to note that the PSS during the solar transient event of March 24, which represents a strong geomagnetic event, is of the same order as on March 02, the reference quiet day of the month as shown in the top-left panel of Figure 6. The PSS on March 02 was about -1.65 which represents the magnetic field fluctuations during the period between 04:00 and 08:00 UT.

The PSS values for the other two extreme solar events (i.e., May 10-11, and October 10, 2024) are respectively shown in the bottom left panels of Figure 6. For the May 10-11 event, the magnetic field fluctuation data observed between 17:00 UT on May 10 and 12:00 UT on May 11, 2024 have been considered for the analysis. As shown in Figure 3, magnetic field fluctuation during this period is quite intense. Similarly, for generating PSS values for the October 10 event, between 14:30 and 22:00 UT has been used. It is to be noted that the PSS values on these days are almost of the same order (-1.69 on May 10-11, and -1.66 on October 10) as on March 24, 2024. We may note that although during all the three events the PSS follow the Kolmogorov scaling, the three events were different in nature from each other.

In the following subsection we will discuss what these PSS values tell us about the turbulence in the solar wind plasma and its impact on energetics of the ambient interplanetary medium.

5.1.1 PSS: What does it tell us about turbulence in the solar wind plasma?

On macroscopic scales, space plasmas behave like turbulent fluids influenced by magnetic fields; an interaction governed by the MHD. The solar wind, which is a continuous outflow of charged particles from the Sun, serves as a natural laboratory for studying plasma turbulence. Similarly, the near-Earth plasma environment at 1 AU exhibits a high magnetic Reynolds number, indicating a highly turbulent state (Kiyani et al., 2015). This turbulence is characterized by a PSS that reveals an energy cascade, where energy from large-scale fluctuations is transferred progressively to smaller scales until it is ultimately dissipated. It is a measure for the nature of fluctuations at different scales - from the nature of energy injection or driving, to cascade and kinetic dissipation (Horbury et al., 2008; Bandyopadhyay et al., 2021).

The near-Earth solar wind has very low particle densities, ≈ 5 particles cm$^{-3}$ near 1 AU. This low density implies infrequent collisions between particles, classifying the solar wind as a collisionless plasma, where the mean free path can be 1 AU. Consequently, classical viscous dissipation mechanisms are ineffective. Similarly, Joule heating due to electrical resistivity is negligible because the solar wind is highly conductive. These conditions suggest that energy dissipation primarily occurs through interactions between plasma particles and electromagnetic fields. Among such processes, magnetic reconnection, where magnetic energy is rapidly converted into kinetic and thermal energy of particles, emerges as a key dissipation mechanism.

Solar wind turbulence spans a vast range of spatial scales, from millions of kilometres down to electron gyro-radii, which are less than a kilometre. The PSD of magnetic field fluctuations (MFF) in the solar wind at 1 AU typically exhibits three distinct regions:

1. Source range: At the lowest frequencies, corresponding to observational periods of several days, the PSD follows an $f^{-1}$ scaling (Kiyani et al., 2015). The correlation length in this region is around $10^6$ km. At slightly higher frequencies, the MFF no longer retain signatures of their solar origin, reflecting instead the *in situ* dynamics of the solar wind as it propagates through the heliosphere. Particularly around the regions near Sun, the solar wind falls in the source range (Goldstein, 2001).
2. Inertial range: This is the region where the classical fluid energy cascade occurs, transferring energy from large to smaller scales. At MHD scales, the PSD follows a Kolmogorov-like power-law with a slope close to $f^{-5/3}$ (PSS ~ -1.67). The dynamics in this range are dominated by the energy injection rate and the nonlinear interaction of turbulent eddies.
3. Dissipation range: Located between the ion and electron gyro-scales, this range marks a transition where the MHD description breaks down. Here, the power-law steepens due to kinetic effects of individual particles. The slope varies depending on local plasma and magnetic conditions, reflecting the nature of dissipation where electromagnetic energy is transferred to particles via kinetic mechanisms.

A representative plot of the PSD of magnetic field fluctuations in the solar wind at 1 AU is shown in Figure 7. Notably, while the PSD generally follows the Kolmogorov power-law scaling in the inertial range during extreme solar events, a significant deviation is observed during the solar quiet period on March 02. During this interval, the spectral slope steepens to approximately -1.83; substantially steeper than during more disturbed solar wind conditions. This suggests that, in addition to the classical picture of a fully developed turbulent cascade, where energy transfers from large to small eddies, other physical mechanisms influence the turbulence dynamics.

The standard Kolmogorov framework assumes turbulence to be statistically steady, homogeneous, and isotropic. However, these assumptions are often violated in the solar wind, a highly anisotropic and inhomogeneous plasma. A key source of anisotropy is the large-scale mean magnetic field, which constrains fluctuations and results in different scaling laws parallel and perpendicular to the field direction (Goldreich & Sridhar, 1995; Horbury et al., 2008). The radial expansion of the solar wind also introduces large-scale gradients that induce scale-dependent anisotropies (Verdini & Grappin, 2015).

Wave–turbulence interactions here play a significant role: Alfvénic fluctuations, common in the solar wind, preferentially propagate along magnetic field lines and suppress nonlinear interactions in that direction (Tu, 1995). Moreover, turbulence is frequently imbalanced, with more energy in outward than inward propagating modes, altering the cascade dynamics (Lithwick, 2007). At smaller scales, kinetic effects and temperature anisotropies can lead to micro-instabilities that further modify magnetic fluctuations (Sahraoui, 2009, Chen, 2016).

These complexities give rise to an intermittent, non-uniform cascade dominated by structures such as current sheets and vortex tubes. These features become increasingly sparse at smaller scales, leading to spectral steepening and deviations from Kolmogorov's prediction. The stretching of vortex lines, associated with vortex tubes, enhances vorticity in three-dimensional flows, but this requires some viscosity for sustained dynamics, a factor not captured by ideal MHD models.

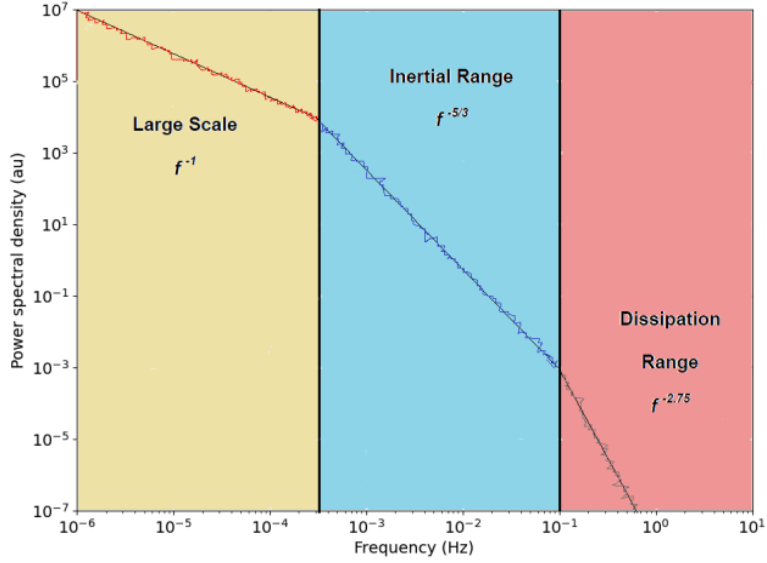

Figure 7: Schematic diagram describing PSD of magnetic field fluctuations.

5.2 Anisotropy in the Magnetic field fluctuations during the strong geomagnetic storm events

To investigate anisotropy in the IMF turbulence, we analyzed the PSS of magnetic field fluctuations along the three IMF axes at the L1 point on two contrasting days: March 24, 2024, marked by a strong geomagnetic storm (Dst ~ –120 nT), and March 02, 2024, a geomagnetically quiet interval (Dst ~ –10 nT), as shown in Figure 8.

On March 24, the PSS for all components fall within the range –1.64 to –1.76, consistent with the Kolmogorov inertial-range value of (-5/3), indicating fully developed, isotropic turbulence. The similarity in spectral slopes across the IMF components supports the interpretation of an energy cascade that is largely direction-independent under disturbed solar wind conditions.

Conversely, on March 02, a marked deviation is observed in the Bz component, which exhibits a steeper spectral index (-1.83) suggestive of a transitional regime between the inertial and dissipation scales.

The increased variability and slope steepening are indicative of developing anisotropy in the magnetic field turbulence, potentially governed by critical balance dynamics, where the anisotropic cascade follows (Goldreich & Sridhar, 1995):

$$k_{\parallel} \propto k_{\perp}^{2/3}$$

Figure 8 presents analogous PSS measurements for two additional, more intense solar events. The upper panels of this Figure show PSS along the three components of IMF during May 10-11, 2024 event, and the lower panels show the same but for October 10, 2024 solar extreme event. It is worth recalling that the May 10-11 event was a "great" geomagnetic event with Dst ~ -410 nT (Figure 3), while the October 10 event was a severe geomagnetic storm event having Dst value of ~ -250 nT (Figure 4). Despite the higher activity levels, the spectral indices on these days closely follow the canonical Kolmogorov value of α ≈ 5/3 for all three components, indicating quasi-isotropic turbulence.

These findings are consistent with the scenario proposed by (Bruno & Carbone, 2013), where large-scale compressive structures associated with strong solar transients tend to suppress directional asymmetries, thereby homogenizing the energy cascade. In contrast, during quiet solar wind conditions, the lack of such large-scale drivers allows inherent anisotropies to develop in the IMF turbulence Matthaeus & Velli, 2011).

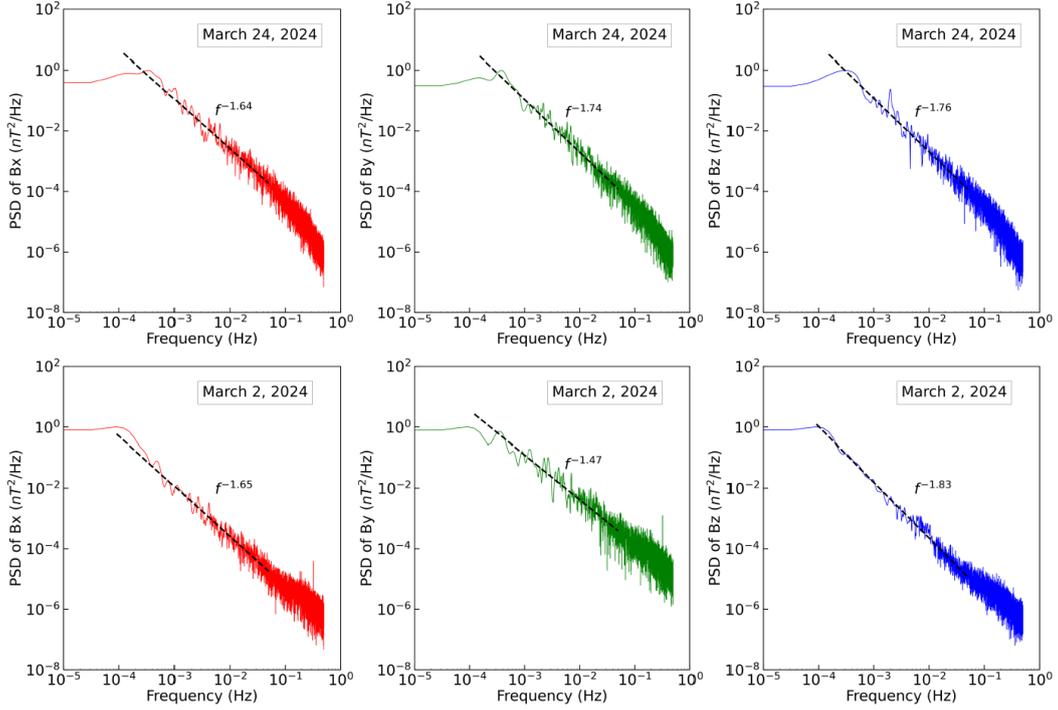

Figure 8: Anisotropy in the IMF during the storm event in March 2024.

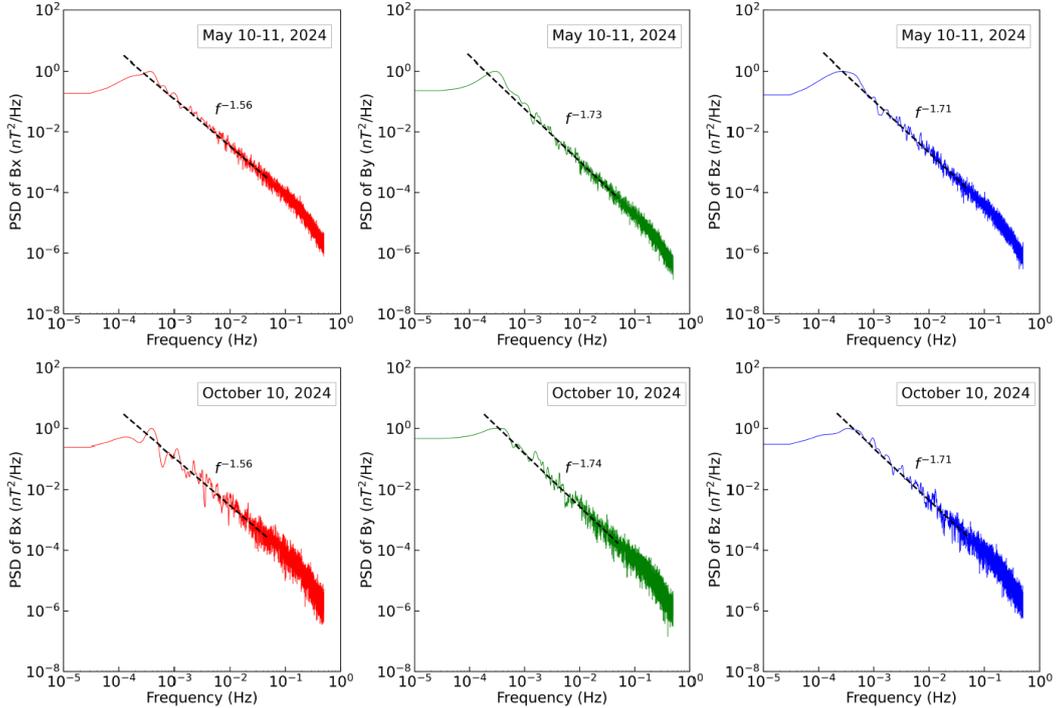

Figure 9: Anisotropy in the IMF during the extreme storm event on May 2024 (upper panel), and October 2024 (lower panels).

In the solar wind, turbulence is often influenced by the presence of a large-scale IMF, which tends to induce anisotropy. However, under certain conditions, particularly in the absence of strong mean fields or in highly turbulent regions where local fluctuations dominate - the magnetic field fluctuations can exhibit near-isotropic behavior.

Studies (Matthaeus et al., 1990; Horbury et al., 2008; and Chen et al., 2010) have explored the scale-dependent nature of anisotropy in solar wind turbulence and its possible transition toward isotropy under certain plasma conditions. An isotropic behavior may emerge under the following conditions:

i. At kinetic scales (i.e., ion and electron gyro-scales), where the turbulent cascade becomes increasingly decoupled from the global $\mathbf{B_0}$, and instead governed by local field variations.
ii. In high plasma-β environments, where thermal pressure dominates over magnetic pressure, diminishing the influence of the background magnetic field.
iii. In the 2D turbulence regime, where fluctuations predominantly lie in the perpendicular plane and exhibit rotational symmetry within that plane.

(Chen et al., 2010), using multi-spacecraft measurements from the *Cluster* mission, demonstrated that the spectral energy distribution of fluctuations becomes more isotropic at sub-ion scales.

At frequencies below the local proton gyrofrequency, where Alfvén waves propagate, fluctuations of the magnetic field are observed with the individual components fluctuating much more than the total magnetic field (|B|). Such small-scale turbulent fluctuations in the magnetic field are of greatest amplitude in the fast wind with the magnetic field fluctuations being the order of magnitude as the average field itself.

## 6. Concluding Remarks

The IMF data, obtained from the MAG payload onboard *Aditya-L1* spacecraft, operating in a halo orbit around L1 point, is analyzed for magnetic fluctuation studies. In 2024, the MAG instrument recorded multiple solar transient events, out of which three extreme events, occurring in March, May, and October, are selected for detailed investigation of magnetic field turbulence in the IMF.

Each of these events exhibited distinct characteristics in terms of IMF amplitude enhancements and temporal duration. Notably, the event from May 10-12, 2024 was the most intense, associated with a "great" geomagnetic storm (Dst < -400 nT), and featured magnetic field fluctuations that significantly exceeded the mean IMF values. For each event, we computed the power spectral slope of the IMF components. Remarkably, all three events exhibited spectral indices close to the Kolmogorov inertial-range value of (-5/3) which is characteristic of fully developed MHD turbulence (Kolmogorov, 1991, Bruno, 2013).

To examine contrast in turbulent scaling under quiescent solar conditions, we compared the PSS values during these events with those observed on March 02, 2024, a geomagnetically quiet day (Dst ≈ –10~nT). Interestingly, the quiet day displayed a spectral slope (~ -1.65) in the total |B| value, which is of the same order as on the geomagnetically disturbed days. A temporal analysis of the PSS evolution during each CME event also revealed a consistent trend: spectral slopes tended to peak near the maximum IMF amplitude, implying enhanced energy transfer during the core phase of each storm. However, while the storm periods displayed quasi-isotropic scaling, evidenced by similar spectral indices across the three IMF components, the quiet day exhibited marked anisotropy. The PSS on March 02, 2024 ranged from –1.47 to –1.83 across components, with the southward Bz component showing the steepest slope, indicative of a more active turbulent cascade in that direction. This suggests that during quiet intervals, the lack of large-scale drivers may allow inherent anisotropic structures to dominate the energy cascade.

These results underscore the variability of IMF turbulence under different solar wind driving conditions and point to the need for deeper theoretical investigation into the interplay between isotropic and anisotropic turbulence regimes in both active and quiescent solar wind environments.

## 7. Acknowledgments

The valuable inputs provided by the reviewers and the suggested references were deeply valued, leading to significant enhancements in the article's quality. This work is supported by Indian Space Research Organization (ISRO). The authors extend their sincere gratitude to the MAG instrument team at LEOS, Bengaluru; The Mechanism and Thermal teams at URSC, Bengaluru for realizing the MAG payload and Srikar P.T., SAG, URSC, Bengaluru for his help and support for MAG ground segment and data pipeline. The authors also express their appreciation and gratitude to all the members of various teams of the Aditya-L1 mission - the Mission and Flight Dynamics Group (FDG) team for helping in the L0 data generation including the required HK and SPICE data of Aditya-L1; the Indian Space Science Data Centre (ISSDC) and ISRO Satellite Tracking centre (ISTRAC) teams for data downloading from the spacecraft and communication with the Payload Operation Centre (POC); the Mission Planning and Operations team, etc. The authors also acknowledge the useful scientific discussions with Dr. Tarun K. Pant (Director, SPL/VSSC, Thiruvananthapuram, India) and Mr. Keshav Aggarwal (DAASE IIT, Indore, India) during this work.

## References

Aggarwal, K., Choudhary, R. K., Datta, A., Mv, R., & Dai, B. K., 2025, Insights into Solar Wind Flow Speeds from the Coronal Radio Occultation Experiment: Findings from the Indian Mars Orbiter Mission, The Astrophysical Journal, 982, 152, doi:10.3847/1538-4357/adb627

Ambili, K. M., Choudhary, R. K., Ashok, A., et al., 2025, The impact of 11 May 2024 great geomagnetic storm on the plasma distribution over the Indian equatorial/low latitude ionospheric region, Journal of Geophysical Research: Space Physics, 130, e2024JA033368, doi:10.1029/2024JA033368

Bandyopadhyay, R., McComas, D. J., Szalay, J. R., et al., 2021, Observation of Kolmogorov turbulence in the Jovian magnetosheath from JADE data, Geophysical Research Letters, 48, doi:10.1029/2021GL095006

Bruno, R., & Carbone, V., 2013, The solar wind as a turbulence laboratory, Living Reviews in Solar Physics, 10, 2, doi:10.12942/lrsp-2013-2

Bruno, R., et al., 2007, Intermittent character of interplanetary magnetic field fluctuations, Physics of Plasmas, 14, 032901, doi:10.1063/1.2711429

Chen, C.H.K. 2016, Recent progress in astrophysical plasma turbulence from solar wind observations, Journal of Plasma Physics, 82, 535820602, doi:10.1017/S0022377816001124

Chen, C.H.K., Horbury, T. S., Schekochihin, A. A., et al., 2010, Anisotropy of solar wind turbulence between ion and electron scales, Phys. Rev. Lett., 104, 255002, doi:10.1103/PhysRevLett.104.255002

Couzens, D. A., & King, J. H., 1986, Interplanetary Medium Data Book-Supplement 3, National Space Science Data Center, Goddard Space Flight Center, Greenbelt, Maryland, USA

Efimov, A. I., Samoznaev, L. N., Bird, M. K., IV, C., & Plettemeier, D., 2008, Solar wind turbulence during the solar cycle deduced from Galileo coronal radio-sounding experiments, Advances in Space Research, 42, 117, doi:10.1016/j.asr.2008.03.025

Garrett, H., 1974, The role of fluctuations in the interplanetary magnetic field in determining the magnitude of substorm activity, Planetary and Space Science, 22, 111, doi:10.1016/0032-0633(74)90127-5

Goldreich, P., & Sridhar, S., 1995, Toward a theory of interstellar turbulence 2: Strong Alfvénic turbulence, Astrophysical Journal, 438, 763, doi:10.1086/175121

Goldstein, M.L., 2001, Major unsolved problems in space plasma physics, Astrophysics and Space Science, 277, 349, doi:10.1023/A:1012264131485

Good, S. W., Ala-Lahti, M., Palmerio, E., Kilpua, E. K. J., & Osmane, A., 2020, Radial Evolution of Magnetic Field Fluctuations in an Interplanetary Coronal Mass Ejection Sheath, The Astrophysical Journal, 893, 110, doi:10.3847/1538-4357/ab7fa2

Goyal, S. K., Tiwari, N. K., Patel, A. R., et al., 2025, Aditya Solar Wind Particle Experiment on Board Aditya--L1: The Supra-Thermal and Energetic Particle Spectrometer, Solar Physics, 300, 35, doi:10.1007/s11207-025-02441-z

Hairston, M. R., & Heelis, R.A., 1990, Model of the high-latitude ionospheric convection pattern during southward interplanetary magnetic field using DE 2 data, Journal of Geophysical Research: Space Physics, 95, 2333, doi:10.1029/JA095iA03p02333

Hapgood, M. A., Lockwood, M., Bowe, G. A., Willis, D. M., & Tulunay, Y. K., 1991, Variability of the interplanetary medium at 1 AU over 24 years: 1963-1986, Planetary and Space Science, 39, 411, doi:10.1016/0032-0633(91)90003-S

Hayakawa, H., Ebihara, Y., Mishev, A., et al., 2025, The Solar and Geomagnetic Storms in 2024 May: A Flash Data Report, The Astrophysical Journal, 979, 49, doi:10.3847/1538-4357/ad9335

Horbury, T. S., Forman, M., & Oughton, S., 2008, Anisotropic scaling of magnetohydrodynamic turbulence, Phys. Rev. Lett., 101, 175005, doi:10.1103/PhysRevLett.101.175005

Hu, Q., Zank, G. P., Li, G., & Ao, X., 2013, A power spectral analysis of turbulence associated with interplanetary shock waves, AIP Conference Proceedings, 1539, 175, doi:10.1063/1.4811016

Jain, A., Trivedi, R., Jain, S., & Choudhary, R. K., 2025, Effects of the Super Intense Geomagnetic Storm on 10-11 May, 2024 on Total Electron Content at Bhopal, Advances in Space Research, 75, 953, doi:10.1016/j.asr.2024.09.029

Jain, R. N., Choudhary, R. K., Bhardwaj, A., et al., 2023, Turbulence dynamics and flow speeds in the inner solar corona: results from radio-sounding experiments by the Akatsuki spacecraft, Monthly Notices of the Royal Astronomical Society, 525, 3730, doi:10.1093/mnras/stad2491

Jain, R. N., Choudhary, R. K., & Imamura, T., 2024, Exploring the influence of the 'Smiley Sun' on the dynamics of inner solar corona and near-Earth space environment, Monthly Notices of the Royal Astronomical Society: Letters, 529, L123, doi:10.1093/mnrasl/slae008

Janardhan, P., Vadawale, S., Bapat, B., et al., 2017, Probing the heliosphere using in-situ payloads on-board Aditya-L1, Current Science, 113, 620, doi:10.18520/cs/v113/i04/620-624

Kasper, J. C., Klein, K. G., Lichko, E., et al., 2021, Parker Solar Probe Enters the Magnetically Dominated Solar Corona, Phys. Rev. Lett., 127, 255101, doi:10.1103/PhysRevLett.127.255101

Kawashima, N., 1969, Analysis of fluctuations in the interplanetary magnetic field obtained by IMP 2, Journal of Geophysical Research (1896-1977), 74, 225, doi:10.1029/JA074i001p00225

Kiyani, K. H., Osman, K. T., & Chapman, S. C., 2015, Dissipation and heating in solar wind turbulence: From the macro to the micro and back again, Philos. Trans. R. Soc. A, 373, 20140155, doi:10.1098/rsta.2014.0155

Kolmogorov, A. N., 1991, The Local Structure of Turbulence in Incompressible Viscous Fluid for Very Large Reynolds Numbers, Proceedings of the Royal Society of London Series A, 434, 9, doi:10.1098/rspa.1991.0075

Kumar, P., Bapat, B., Shah, M. S., et al., 2025, Aditya Solar Wind Particle Experiment (ASPEX) on Board Aditya-L1: The Solar Wind Ion Spectrometer (SWIS), Solar Physics, 300, 37, doi:10.1007/s11207-025-02443-x

Lee, K. H., & Lee, L. C., 2020, Turbulence Spectra of Electron Density and Magnetic Field Fluctuations in the Local Interstellar Medium, The Astrophysical Journal, 904, 66, doi:10.3847/1538-4357/abba20

Lithwick, Y., Goldreich, P., & Sridhar, S., 2007, Imbalanced strong MHD turbulence, The Astrophysical Journal, 655, 269, doi:10.1086/509884

Loewe, C. A., & Prölss, G. W., 1997, Classification and mean behavior of magnetic storms, Journal of Geophysical Research: Space Physics, 102, 14209, doi:10.1029/96JA04020

Matthaeus, W. H., Goldstein, M. L., & Roberts, D.A., 1990, Evidence for the presence of quasi-two-dimensional nearly incompressible fluctuations in the solar wind, Journal of Geophysical Research: Space Physics, 95, 20673, doi:10.1029/JA095iA12p20673

Matthaeus, W. H., & Velli, M., 2011, Who needs turbulence? A review of turbulence effects in the heliosphere and on the fundamental process of reconnection, Space Science Reviews, 160, 145, doi:10.1007/s11214-011-9793-9

Moissard, C., Fontaine, D., & Savoini, P., 2019, A Study of Fluctuations in Magnetic Cloud-Driven Sheaths, Journal of Geophysical Research: Space Physics, 124, 8208, doi:10.1029/2019JA026952

Pi, G., Pitňa, A., Zhao, G.-Q., et al., 2022, Properties of Magnetic Field Fluctuations in Long-Lasting Radial IMF Events from Wind Observation, Atmosphere, 13, 173, doi:10.3390/atmos13020173

Ram, S. T., Veenadhari, B., Dimri, A. P., et al., 2024, Super-intense geomagnetic storm on 10-11 May 2024: Possible mechanisms and impacts, Space Weather, 22, e2024SW004126, doi:10.1029/2024SW004126

Sahraoui, F., Goldstein, M. L., Robert, P. & Khotyaintsev, Y. V., 2009, Evidence of a cascade and dissipation of solar-wind turbulence at the electron gyroscale, Phys. Rev. Lett., 102, 231102, doi:10.1103/PhysRevLett.102.231102

Sankarasubramanian, K., Bug, M., Sarvade, A., et al., 2025, Solar Low Energy X-ray Spectrometer (SoLEXS) on Board Aditya-L1 Mission, Solar Physics, 300, 87, doi: 10.1007/s11207-025-02494-0

Sheeley, N. R. & Wang, Y. -M., 2003, On the fluctuating component of the Sun's large-scale magnetic field, The Astrophysical Journal, 590, 1111, doi:10.1086/375026

Singh J., Ramesh, R., Prasad, B. R., et al., 2025, Visible Emission Line Coronagraph (VELC) on Board Aditya-L1, Solar Physics, 300, 66, doi:10.1007/s11207-025-02477-1

Smith, C. W., Vasquez, B. J. & Hamilton, K., 2006, Interplanetary magnetic fluctuation anisotropy in the inertial range, Journal of Geophysical Research: Space Physics, 111, A09111, doi: 10.1029/2006JA011651

Smith, M. F., & Lockwood, M., 1996, Earth's magnetospheric cusps, Reviews of Geophysics, 34, 233, doi:10.1029/96RG00893

Stamper, R., Lockwood, M., Wild, M. N., & Clark, T. D. G. 1999, Solar causes of the long-term increase in geomagnetic activity, Journal of Geophysical Research: Space Physics, 104, 28325, doi:10.1029/1999JA900311

Thampi, R. S., Abhishek, J. K., Sasidharan, D., et al., 2025, Plasma-Analyzer Package for Aditya (PAPA) on Board the Indian Aditya-L1 Mission, Solar Physics, 300, 9, doi:10.1007/s11207-024-02414-8

Tripathi, D., Chakrabarty, D., Nandi, A., et. al., 2022, The Aditya-L1 mission of ISRO, Proceedings of the International Astronomical Union, 18, 17–27, doi:10.1017/S1743921323001230

Tripathi, D., Ramaprakash, A. N., Padinhatteeri, S., et al., 2025,The Solar Ultraviolet Imaging Telescope on Board Aditya-L1, Solar Physics, 300, 30, doi:10.1007/s11207-025-02423-1

Tu, C.-Y., & Marsch, E., 1995, MHD structures, waves and turbulence in the solar wind: Observations and theories, Space Science Reviews, 73, 1, doi:10.1007/BF00748891

Verdini, A., & Grappin, R., 2015, Imprints of expansion on the local anisotropy of solar wind turbulence, The Astrophysical Journal Letters, 808, L34, doi:10.1088/2041-8205/808/2/L34

Wang, Y.-M., & Sheeley, N. R., 1988, The solar origin of long-term variations of the interplanetary magnetic field strength, Journal of Geophysical Research: Space Physics, 93, 11227, doi:10.1029/JA093iA10p11227

Welch, P., 2003, The use of fast Fourier transform for the estimation of power spectra: A method based on time averaging over short, modified periodograms, IEEE Transactions on audio and electroacoustics, 15, 70

Xu, S. B., Huang, S. Y., Sahraoui, F., et al., 2023, Observations of Kolmogorov Turbulence in Saturn's Magnetosphere, Geophysical Research Letters, 50, e2023GL105463, doi:10.1029/2023GL105463

Yadav, V. K., Srivastava, N., Ghosh, S., Srikar, P. T., & Subhalakshmi, K., 2018, Science objectives of the magnetic field experiment onboard Aditya-L1 spacecraft, Advances in Space Research, 61, 749, doi: 10.1016/j.asr.2017.11.008

Yadav, V. K., Vijaya, Y., Krishnam Prasad, B., et al., 2025, The Fluxgate Magnetometer (MAG) on board Aditya-L1 Spacecraft, Solar Physics, 300, 33, doi: 10.1007/s11207-025-02440-0

Zhao X. & Hoeksema, J. T, 1995, Prediction of the interplanetary magnetic field strength, Journal of Geophysical Research: Space Physics, 100, 19, doi:10.1029/94JA02266